\documentclass[aps,pre,floats,floatfix,twocolumn]{revtex4}

\usepackage{ifthen}
\usepackage{ifpdf}
\usepackage{color}
\usepackage{graphicx}
\usepackage{epsfig}
\graphicspath{{./Figs/}{./}}

\usepackage{latexsym}
\usepackage{amsmath}
\usepackage{amssymb}
\usepackage{bm}
\usepackage{wasysym}

\newcommand{\eexp}{\mbox{e}^}

\newcommand{\mylabel}[1]{\label{#1}}  
\newcommand{\beq}{\begin{eqnarray}}
\newcommand{\eeq}{\end{eqnarray}} 
\newcommand{\be}[1]{\begin{eqnarray}\ifthenelse{#1=-1}{\nonumber}{\ifthenelse{#1=0}{}{\mylabel{e#1}}}}
\newcommand{\ee}{\end{eqnarray}} 

\newcommand{\Eq}[1]{\textcolor{blue}{Eq.\!\!~(\ref{#1})}} 
\newcommand{\Fig}[1]{\textcolor{blue}{Fig.}\!\!~\ref{#1}}
\newcommand{\hide}[1]{\textcolor{red}{[hidden text]}} 

\newcommand{\sect}[1]{{\bf #1.-- }}
\renewcommand{\cite}[1]{\textcolor{blue}{[\onlinecite{#1}}]} 

\newcommand{\ola}{\protect\overleftarrow}
\newcommand{\ora}{\protect\overrightarrow}

\begin{document}

\title{Non-equilibrium version of the Einstein relation} 

\author{Daniel Hurowitz, Doron Cohen}

\affiliation{
\mbox{Department of Physics, Ben-Gurion University of the Negev, Beer-Sheva 84105, Israel}
}

\begin{abstract}
The celebrated Einstein relation between the diffusion coefficient $D$ and the 
drift velocity $v$ is violated in non-equilibrium circumstances. 
We analyze how this violation emerges for the simplest example 
of a Brownian motion on a lattice, taking into account the interplay 
between the periodicity, the randomness and the asymmetry of the transition rates. 
Based on the non-equilibrium fluctuation theorem the $v/D$ ratio is found to be 
a non-linear function of the affinity. Hence it depends in a non-trivial way 
on the microscopics of the sample.
\end{abstract}

\maketitle

\section{Introduction}

The Einstein-Smoluchowski relation (ESR) \cite{ESR1,ESR2} between 
the diffusion coefficient ($D$) and the mobility ($\mu$) 
of a Brownian particle is a landmark in the history of statistical mechanics.
It states that $D = \mu k_B T$, where $T$ is the temperature, 
and $k_B$ is the Boltzmann constant. Thus it reflects the 
microscopics of the stochastic process (via $k_B$) in a very universal way.
Below we set ${k_B=1}$.

The ESR constitues the simplest example for a fluctuation-disspation relation.
In a modern perspective it can be regarded as a consequence 
of a general non-equilibrium fluctuation theorem (NFT) \cite{eprd1,eprd2,eprd3,eprd4}
that concerns the evolving probability distribution $p(x;t)$ 
that describes the stochastic motion of the particle.  
The ESR is in essence a relation between the second moment of 
the spreading $\text{Var}(x)=2Dt$, and its first moment $\langle x \rangle=vt$, 
where ${v=\mu F}$ is the drift velocity, and $F$ is the field of force.
Using this language it can be re-written as follows:  
\be{1}
\frac{v}{D} \ \ = \ \ f_{\sigma}(s)
\ee
where $s=F/T$ {is the so-called {\em affinity} in units of} 
entropy production per unit distance, 
and $f_{\sigma}(s)=s$ is a universal function that does not 
depend on the microscopic details of the sample.

\sect{Model of interest}
We shall consider below the dynamics of a particle on an $N$~site ring, 
with transition rates $\ora{w}_n$ and $\ola{w}_n$ 
across the $n^{th}$ bond. In general the transition rates 
are random and asymmetric. In previous publications~\cite{ner,nef}
we have highlighted the relevance of Sinai spreading~\cite{sinai} 
to the analysis of the induced~$v$.   
Optionally one may have in mind the unfolded version of our ring.  
The latter concerns the motion of a Brownian particle 
in a tilted periodic array of identically disordered unit cells.

\sect{Previous studies}
The dramatic influence of a tilt on the transport in a one-dimensional 
biased {\em periodic} potential has been explored experimentally 
for a colloidal particle on a corrugated optical vortex \cite{tiltexp1,tiltexp2}, 
and has been exploited for optical particle fractionation and separation \cite{srt1,srt2,srt3}.   
Explicit expressions for $v$ and for $D$ for the case of a tilted cosine potential 
were first given in \cite{tilt0,tilt1,tilt2,tilt22} and further generalised in \cite{tilt3,tilt4,tilt5,tilt6}.

Several works have studied the effect of weak spatial {\em disorder}
on the non-linear bias dependence of the transport coefficients \cite{disorder1,disorder2,disorder3}.
Tractable expressions for $v$ and for $D$ were available 
for a completely disordered lattice ($N=\infty$) \cite{odh1,odh2}.
The prediction is that for small~$s$ one obtains~${v=0}$.  
This anomaly is related to the work of Sinai \cite{sinai} 
regarding random walk in random environment. 

Strangely enough there was no attempt, as far as we know, 
to bridge between the implied~$v/D$ dependence on~$s$,  
and the ESR that is expected close to equilibrium. 
Furthermore there was no  attempt to settle what looks like a contradiction 
with the NFT-based derivation of the ESR, which relies on the central limit theorem. 
It is the purpose of the present work to illuminate the departure from the ESR, 
and to explore the route that leads to the $N=\infty$ Sinai anomaly.

For completeness we note that extensions of the fluctuation-dissipation phenomenology 
far from equilibrium have been considered in \cite{udo0,udo,udo2,kbb}, 
but from a different perspective. 
In \cite{udo3} it has been pointed out that a violation of the ESR 
is expected in a Markovian network, however this has not been explicitly 
demonstrated for a model of interest, neither related to the Sinai anomaly.

\section{NFT based derivation}

The NFT relates the probability of a stochastic trajectory $\bm{r}(t)$
to the probability of the time reversed process: 
\be{104}
\frac{P\left[\bm{r}(-t)\right]}{P\left[\bm{r}(t)\right]} \ = \ \exp \left[ -\mathcal{S}[\bm{r}] \right] 
\eeq
where $\mathcal{S}[\bm{r}]$ is the entropy production that is associated 
with the trajectory. 
The implicit assumption here is that $\mathcal{S}[\bm{r}]$ is well defined. 
This is a very strong assumption because in an actual experiment $\mathcal{S}$ 
might depend on additional "hidden" microscopic coordinates that cannot 
be resolved by the measuring device. This can be summarized by saying that 
coarse-graining might make the "bare" NFT inapplicable: possibly 
an effective version of $\mathcal{S}[\bm{r}]$ should be defined \cite{saar,gsprd1,gsprd2}.  
In our model $\mathcal{S}[\bm{r}]$ is a well defined functional, still we shall 
see later that in some sense~$s$ is renormalized due to coarse-graining.   

We proceed with a critical overview of the the derivation of the traditional ESR 
based on \Eq{e104}. The entropy production during one trip around the ring is
\be{105}
\mathcal{S}_{\circlearrowleft}  \ \ = \ \ 
\sum_{n\in\text{ring}} \ \ln \left[ \ \frac{\ora{w}_n}{\ola{w}_n} \ \right ] 
\eeq
{The terms in this sum can be regarded as a {\em stochastic field}.
Summing over the terms along the closed loop we get the {\em stochastic motive force (SMF)}. }
In the context of molecular motors the SMF is known as the {\em affinity}. 
The entropy production for a general trajectory that has 
a winding number~$q$ is  ${\mathcal{S}[r]=q \mathcal{S}_{\circlearrowleft}}$. 
We disregard here small fractional-loop error that can be neglected in the infinite 
time limit. We define formally the distance as ${x=qN}$, 
and the  entropy per unit distance as ${s=\mathcal{S}_{\circlearrowleft}/N}$. 
If follows from \Eq{e104} that the evolving probability distribution satisfies  
\be{108}
\frac{p(-x;t)}{p(x;t)} \ \ = \ \ \eexp{- s x}
\ee
In the long time limit, by virtue of the central limit theorem (CLT), 
one can introduce a Gaussian approximation $p(x) \approx \overline{p}(x)$, 
where 
\beq
\overline{p}(x;t) \ \ = \ \ 
\frac{1}{\sqrt{4\pi Dt}} 
\exp
\left[ 
-\frac{\left(x - v t \right)^2}{4Dt}
\right]
\eeq
Substitution in \Eq{e108} leads to the standard ESR, namely ${v/D=s}$.  
Below we are going to argue that the last step should be handled with 
much more care. The coarse-grained distribution $\overline{p}(x)$ 
that appears in the CLT, is in fact a convoluted ("smoothed") version 
of the bare $p(x)$. If follows that $\overline{p}(x)$ obeys 
a "dressed" version of \Eq{e108}, 
with effective affinity $\overline{s}=f_{\sigma}(s)$, where 
\be{2}
f_{\sigma}(s) \ \ = \ \ \frac{2}{a_s} \tanh\left(\frac{a_ss}{2}\right)
\eeq       
The length scale $a_s$ is related to the 
microscopic details of the model. We first clarify
this statement for the simplest case of a non-disordered 
ring, for which $a_s=1$ is the lattice constant, 
and later discuss the general case. 

{
At this point one should realize that \Eq{e1} with \Eq{e2} 
can be regarded as a generalized ESR. The $v/D$ ratio 
is no longer a universal function because there is 
an affinity dependent length-scale $a_s$, that reflects  
the microscopic structure. The standard ESR is recovered  
if  ${a_s s \ll 1}$. In the other extreme we get the simple relation 
\be{91}
D \ \ = \ \ \frac{1}{2} v_s a_s
\eeq
where the subscript $s$ has been added in order 
to emphasize the crucial dependence on the affinity.
We would like to illuminate how $a_s$ depends 
on the periodicity $N$, on the strength of the disorder $\sigma$, 
and on the affinity~$s$. }

\section{Effective Affinity}
\label{sect4}

Consider the simplest discrete model with asymmetry. 
All the bonds are identical; the transition rates from left to right are $\ora{w}$,    
and the transition rates from right to left are $\ola{w}$. 
Hence it follows from \Eq{e105} that $\mathcal{S}_{\circlearrowleft} = \ln(\ora{w}/\ola{w})$.  
It is not difficult to find the exact expression 
for the the evolving probability distribution $p(x;t)$. 
The dynamics that is generated by a rate equation can be simulated 
as a random walk process with infinitesimal time steps $\tau$.
The traversed distance is ${x=X_1+...+X_{\mathcal{N}}}$.
The transition probabilities per step are 
\beq
P(X =+1) \ &=& \ p \ \ \equiv \ \ \ora{w}\tau \\
P(X =-1) \ &=& \ q \ \ \equiv \ \ \ola{w}\tau \\
P(X = 0) \ &=& \ 1-p-q 
\eeq
The probability distribution can be obtained from the moment generating function of the process
\beq
Z(k) =  \langle \eexp{-ik x} \rangle  =  \left[p\eexp{-ik} + q\eexp{+ik} + (1-p-q)\right]^{\mathcal{N}}
\eeq
In the continuous time limit $p,q\ll 1$, hence one can expand 
\beq
\ln Z(k)  = \mathcal{N} \left[p \eexp{-ik} + q\eexp{+ik} -(p+q) \right]
+ \mathcal{O}(\mathcal{N}\tau^2)
\eeq
Accordingly, one obtains
\beq
p(x;t) = \int_{-\infty}^{\infty}  \!\! dk \ 
\eexp{ ikx + \left(\ora{w} \eexp{-ik} + \ola{w} \eexp{ik} -(\ola{w}+\ora{w})\right)t}
\eeq
This distribution obviously satisfies the NFT \Eq{e108}, 
which can be easily verified by inverting the sign of the dummy integration variable in $p(-x;t)$ and then shifting
it by a constant  ${k\to k+\ln(\ola{w}/\ora{w})}$. 
Expanding the expression 
in the exponent  in powers of~$k$ one obtains  
\beq
p(x;t) = \int_{-\infty}^{\infty} \!\! dk \
\eexp{ ik (x - (\ora{w}-\ola{w})t) - \frac{k^2}{2}(\ora{w}+\ola{w})t  + \mathcal{O}(k^3 t) }
\eeq
The average $vt$ and the variance $2Dt$ 
are implied by the coefficients 
of the $k$ and $k^2$ terms in the exponent, namely
\be{15}
v \ &=& \ {\left(\ora{w}- \ola{w}\right)} 
\\ \label{e114}
D \ &=& \ \frac{1}{2}\left( \ora{w} + \ola{w}\right)
\eeq
The $v/D$ ratio is given by \Eq{e2} with $a_s=1$.
{This exact result clearly contradicts the traditional ESR.}

{We now would like to see what happens to the NFT and the ESR 
if the CLT is applied.} Recall that the CLT procedure 
is to introduce the re-scaled variable ${(x-vt)/(2Dt)}$ 
and to claim that in the ${t \to \infty}$ limit 
the higher order cumulants $\mathcal{O}(k^3 t)$ can be neglected.
We use the notation $\overline{p}(x;t)$ for the normal distribution 
that is obtained via CLT. {It is easy to see} that it satisfies \Eq{e108}, 
but with an effective value of $s$ that is given by \Eq{e2}.
{Consequently the associated relation ${v/D=\overline{s}}$ is consistent 
with the exact analysis, while the bare relation ${v/D=s}$ is violated.}    

We conclude that the normal approximation for $p(x;t)$,  
which is implied by CLT, obeys the NFT provided $s$ is replaced 
by a renormalized value $\overline{s}$. 
The reason for that is as follows:
The CLT procedure is the same as cutting off the high $k$ modes, 
which is the same as smoothing the function $p(x,t)$. 
Due to the smoothing the effective value of $s$ becomes smaller, 
{and consequently the {\em bare} ESR is violated.}

\section{Drift and Diffusion}

We now turn to describe a general procedure for exact calculation 
of $v$ and $D$. Any rate equation can be written schematically 
as ${d\bm{p}/dt = W\bm{p}}$, where $\bm{p}=\{p_n\}$ is a column vector 
that contains the occupation probabilities, and $W$ is a matrix that contains 
the transition rates. Note that this matrix is non-symmetric, 
hence one should distinguish between right and left eigenvectors.    
If the lattice is periodic, with a unit cell that consists of $N$ sites, 
the eigenevectors satisfy the Bloch theorem. 
The reduced equation for the eigenmodes becomes $\bm{W}(\varphi)\psi = -\lambda\psi$, 
where $\bm{W}(\varphi)$ is an $N\times N$ matrix, and the  
presence of the phase $\varphi$ implies that $\psi_{n+N}=\eexp{i\varphi}\psi_n$,  
where $n$ is the site index mod($N$). The Bloch quasi-momentum 
is formally defined via the relation ${\varphi\equiv kN}$.
The outcome of the diagonalization process are the Bloch state ${|k,\nu\rangle}$, 
where $\nu$ is the band index, and the corresponding eigenvalues 
are $-\lambda_{\nu}(k)$. The bottom line is that the time dependent 
solution of the rate equation can be written as 
\beq
p_n(t) \ \ \approx \ \ \frac{1}{L}\sum_{k,\nu} C_{k,\nu} \ \eexp{-\lambda_{\nu}(k) t} \ \eexp{ikn}
\eeq
where $C_{k,\nu}$ are constants that depend on the initial preparation.
For technical details see {appendix~A}. The long time spreading
is dominated by the lowest band $\nu=0$. It is not difficult 
to show (see appendix) that the drift velocity and the diffusion coefficient 
are determined by the derivatives of~$\lambda_0(k)$. Namely,
\be{116}
v \ &=& \ \left. i \frac{\partial \lambda_0(k)}{\partial k}\right|_{k=0}
\\ \label{e117}
D \ &=& \ \left. \frac{1}{2} \frac{\partial^2 \lambda_0(k)}{\partial k^2}\right|_{k=0}
\eeq

\begin{figure}
\centering
\includegraphics[height=6.5cm]{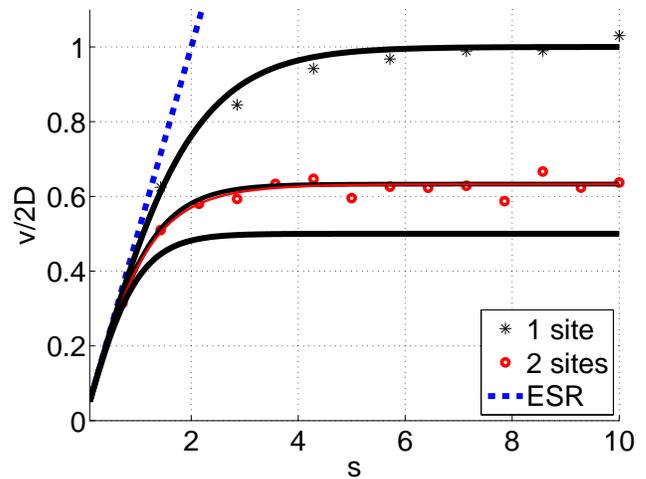}

\caption{ 
The ratio $v/(2D)$ for a Brownian motion in a one-dimensional regular lattice.
The number of sites per unit cell is $N{=}1$ (black stars) and $N{=}2$ (red circles). 
The numerical results (symbols) are based on simulations with ensembles of $10^3$ trajectories, 
while the lines are exact analytical expressions.  
The dashed line is the ESR.
The upper and lower thick solid lines are \Eq{e2} with $a{=}1$ and $a{=}2$ respectively. 
In the $N{=}2$ case ${\ola{A}\ora{A}=\ola{B}\ora{B}=1}$ and ${\sigma=2}$.
The intermediate thick solid line is \Eq{e2} with ${a=a_{\infty}=1.58}$. 
It barely can be resolved from the exact result (thin red line).   
}
\label{fig1}
\end{figure}

\sect{The $N=2$ system}
As an explicit example for the outcome of this procedure 
we consider a periodic lattice 
that has unit cell with $N{=}2$ sites.
The transition rates ${(\ora{A},\ola{A},\ora{B},\ola{B})}$
are characterized by $\ln(\ora{A}/\ola{A})=s{+}\sigma$ 
and ${\ln(\ora{B}/\ola{B})=s{-}\sigma}$, such that~$\sigma$ 
is the ``disorder" (in a later example $\sigma$ will stand for 
the width of a box distribution). 
{After some straightforward algebra (see Appendix~B)} one obtains  
\be{5}
v \ = \ \frac{\ora{A} \ora{B} - \ola{A} \ola{B}}{ \ora{A} + \ola{A} + \ora{B}+  \ola{B}} 
\ee
and
\be{6}
D \ = \ \frac{1}{2}\left[
\frac{\ora{A} \ora{B} + \ola{A} \ola{B}  - 2v^2 }
{ \ora{A}+ \ola{A} + \ora{B}+  \ola{B}}\right] 
\ee
The $v/D$ ratio is given by \Eq{e1} with 
\be{7}
f_{\sigma}(s) \ = \ \frac{2}{1 +  \tanh^2\left(\frac{\sigma}{2}\right)\tanh^2\left(\frac{s}{2}\right)} \tanh\left(\frac{s}{2}\right)
\eeq
This result is compared to a numerical simulation of a random walk in \Fig{fig1}, 
which was obtained by standard simulation methods (Gillespie's algorithm). 
As $s$ is increased the $v/D$ ratio approaches a limiting value 
which we define as ${2/a_{\infty}}$. In \Fig{fig1} we have added a curve  
of the function \Eq{e2} with ${a_s=a_{\infty}}$. We observe that for practical 
purpose $a_{\infty}$ can be regarded as an effective lattice constant. 
As the ``disorder" $\sigma$ increases $\tanh^2({\sigma/2})$ 
grows from~0 to~1, and consequently ${a_{\infty}}$ grows from the 
value ${a=1}$ to the value ${a=2}$. In spite of the simplicity 
of this example we shall see that it provides partial insight with regard 
to the general~$N$ case.

\begin{figure}
\centering
\includegraphics[height=6.5cm]{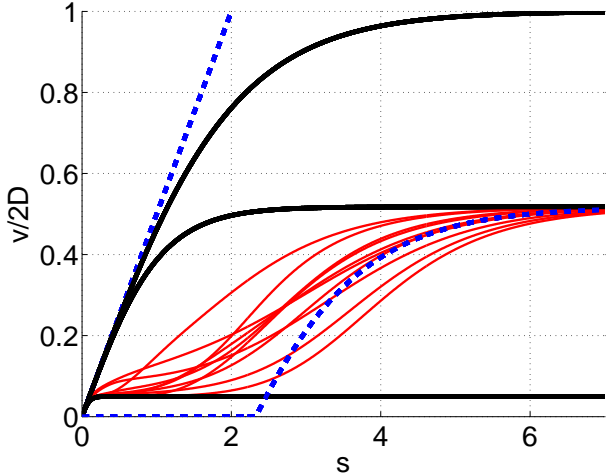}

\caption{
The same as \Fig{fig1} but with $N{=}20$ sites per unit cell.
The upper and lower thick solid lines are for zero disorder ($\sigma{=}0$) 
and for infinite disorder ($\sigma{\to}\infty$), 
as implied by \Eq{e2} with ${a{=}1}$ and ${a{=}20}$ respectively.
The thin solid curves are based on exact analytical calculation 
for various realizations of disorder that is characterized by ${\sigma{=}3.5}$.
The intermediate thick solid line is \Eq{e2} with~$a=a_{\infty}=1.9316$, estimated using \Eq{e3}. 
The linear dashed line is the ESR while the second dashed line that 
exhibits a ``Sinai step" is the $N{=}\infty$ prediction of \cite{odh1} (see text).
}
\label{fig2}
\end{figure}

\sect{General $N$ system}
Let us explore how $f_{\sigma}(s)$ looks like when $N$ becomes 
larger. \Fig{fig2} provides a few examples that were 
calculated analytically using \Eq{e116} and \Eq{e117} for $N{=}20$. 
The rates were chosen as $\ora{w}_n = \eexp{\mathcal{S}_n/2}$ 
and $\ola{w}_n = \eexp{-\mathcal{S}_n/2}$, 
where $\mathcal{S}_n$ are box distributed within ${[s-\sigma,s+\sigma]}$.
This implies that the rates have log-box distribution as in "glassy" systems.

{Looking at the numerical results (\Fig{fig2}), and taking into 
account the simple periodic lattice as a reference case,} 
we observe that $f_{\sigma}(s)$ has some typical properties 
{that we would like to analyze in later sections.} Namely: 
{\bf \ (1)}~For small values of $s$ we have ${f_{\sigma}(s) = s}$ 
in consistency with the ESR. 
{\bf \ (2)}~For no disorder ($\sigma=0$) we already have established 
that $f_{\sigma}(s)$ obeys \Eq{e2} with ${a_s=1}$, 
reflecting the microscopic discreteness of the lattice. 
{\bf \ (3)}~For finite disorder we see that for moderate values 
of $s$ the function $f_{\sigma}(s)$  can be approximated by \Eq{e2} 
with ${a_s=N}$, reflecting the length of the unit cell.
{\bf \ (4)}~For very large values of~$s$ 
the function $f_{\sigma}(s)$ saturates, 
reflecting an effective lattice constant $a_{\infty}$ 
that we discuss {in Section~V}.
{\bf \ (5)}~As $N$ becomes larger our results approach 
those of \cite{odh1}, as discussed {in Section~VII}.

\begin{figure}
\centering
\includegraphics[height=6.5cm]{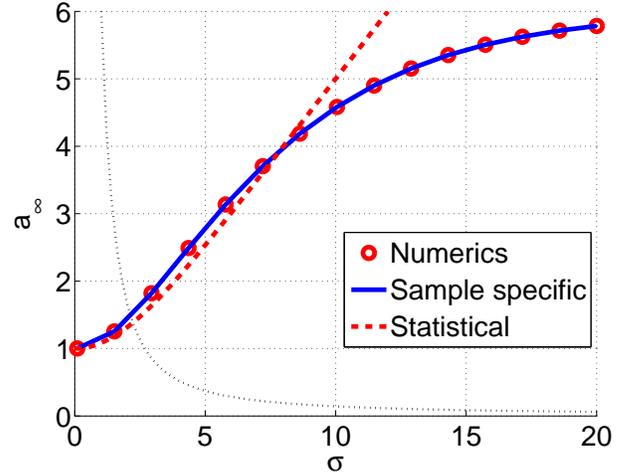}

\caption{
The effective length scale $a_{\infty}=(2D/v)_{s\rightarrow\infty}$. 
The number of sites per unit cell is $N{=}6$.
The symbols are the outcome of an exact analytical calculation. 
The exact sample specific expression \Eq{e3} (solid line) works perfectly, 
independent of~$N$. The statistical approximation \Eq{e123} (dashed line)
becomes indistinguishable for large values of $N$ (not shown).
The dotted line is $a^{(\sigma)}$ of \Eq{e211}.
}
\label{fig3}
\end{figure}

\section{The Poisson limit}

{Going to the extreme of very large $s$ it is possible 
to get simple analytical expressions for $v$ and $D$. 
We first consider a simple periodic lattice. 
From \Eq{e15}-\Eq{e114} we get ${v= \ora{w}a}$
and ${D=(1/2)\ora{w}a^2}$ where $a{=}1$ is the lattice constant.
Accordingly ${v/(2D)=1}$ in consistency with \Eq{e91}.
Let us illuminate the statistical meaning of this result.}   
Recall that the dynamics is generated by a rate equation 
that can be simulated as a random walk process with infinitesimal time steps $\tau$.
The traversed distance is ${x=X_1+...+X_N}$.
For $s\to\infty$ we get a Poisson process 
\beq
P(X = 1) &=& p \\
P(X = 0) &=& 1-p \\
P(X=-1) &=&0 
\eeq
with $p=\ora{w}\tau$. Taking the continuous time limit 
one deduces that in the Poisson limit the ratio between the first 
and the second moment is unity, {hence $v/(2D)$ is determined.}

{We now turn to the disordered case, i.e. the rates are not the same. 
In the Poisson limit we still can get a simple analytical 
expression for $a_{\infty}$, which determines the asymptotic value of $v/(2D)$.}
This is done using the same procedure as in section \ref{sect4}.
The rates are $\ora{w}_n=w_n$ and $\ola{w}_n=0$ 
for ${n=1...N}$. The characteristic equation for the eigenvalues of~$W(\varphi)$ is 
\be{-1}
\det(\lambda+\bm{W}(\varphi)) 
\ = \ \prod_{n=1}^N (\lambda {-} w_n) + \eexp{-i\varphi} \prod_{n=1}^N w_n 
\ = \ 0 
\eeq
This can be re-written as 
\beq
\prod_{n=1}^N \left(1-\frac{\lambda}{w_n} \right) \ \ = \ \ \eexp{-i\varphi} 
\eeq
{Expanding to second order we get}
\beq
\sum_{n=1}^N \frac{\lambda}{w_n} \ - \ \sum_{i\neq j } \frac{\lambda^2}{w_n w_m} 
\ \ = \ \ 1 - \eexp{-i\varphi} 
\ee
with the solution
\beq
\lambda &=& 
-i \left[\left(\sum_{n=1}^N \frac{1}{w_n} \right)^{-1}\right] \varphi 
\\ \nonumber
&+& \frac{1}{2} \left[\left( \sum_{n=1}^N \frac{1}{w_n} \right)^{-3} \left(\sum_{n=1}^N \frac{1}{w_n ^2}\right)\right] \varphi^2 
\ + \ \mathcal{O}(\varphi^3)  
\ee
Using \Eq{e116} and \Eq{e117} we deduce that 
\be{3}
a_{\infty} \ = \  \left(\frac{2D}{v}\right)_{s\rightarrow\infty} \ = \ 
\left[\frac{\big\langle (1/\ora{w})^2 \big\rangle}{\big\langle (1/\ora{w}) \big\rangle^2}\right]
\ee
where the sample average is $\langle R \rangle \equiv (1/N)\sum_n R_n$.
For large $N$ the sample average can replaced 
by an ensemble average.  
Note that the expression in the square brackets constitutes 
a measure for the "glassiness" of the network: 
it becomes much larger than unity due to the presence of weak links.  
For the log-box distributed transition rates of \Fig{fig2} 
\be{123}
a_{\infty} \ \ = \ \ \frac{\sigma}{2} \coth\left( \frac{\sigma}{2} \right)
\eeq
In \Fig{fig3} we test this estimate for $N{=}6$. 
We observe (not shown) that the statistical
result \Eq{e123} becomes indistinguishable from the 
exact sample average \Eq{e3} for large values of~$N$.

\section{Digression - Sinai spreading}

{
In order to understand the dependence of $v$ and $D$ on $N$ and $s$, 
it is useful to recall known results that have been obtained 
for the time-dependent spreading in an $N=\infty$ lattice.
Recall that the drift is induced by the stochastic field, 
whose affinity is defined in \Eq{e105}).  
The comulant generating function of the stochastic field 
can be written as $g(\mu)=(s-s_{\mu})\mu$, where $s_{\mu}$ 
is defined via the following expression:   
\be{362}
\left\langle  \left(\frac{\overleftarrow{w}}{\overrightarrow{w}}\right)^{\mu}\right\rangle \ \ \equiv \ \ \eexp{-(s-s_{\mu})\mu} 
\eeq
If the stochastic field has normal distribution 
with standard deviation $\sigma$, then ${s_{\mu}=(1/2) \sigma^2 \mu}$.
For our box distribution
\beq
s_{\mu}  \ \ = \ \ \frac{1}{\mu} \ln\left( \frac{\sinh (\mu \sigma)}{\mu \sigma} \right)
\eeq
which is monotonically ascending from zero to $\sigma$. 
The positive monotonic function $s_{\mu}$ can be inverted,
hence we can define a scaled affinity $\mu(s)$.
Note that \Eq{e362} implies that $\mu(s)$ is the value 
of $\mu$ for which the left-hand-side equals unity.

The time dependent spreading in the zero bias case ($s{=}0$) 
has been worked out by Sinai (\cite{sinai}),
leading to the anomalous time dependence
\beq
x \ \ \sim \ \ [\log(t)]^2
\eeq
For finite $s$ one obtains \cite{odh3,ohd4}
\beq
x \ \ \sim \ \ t^{\mu}
\eeq
where the exponent $\mu=\mu(s)$ is the ``scaled affinity" 
that has been defined above. 

From the time dependent spreading we can deduce the $N$ 
dependence of the stationary drift velocity. 
The deduction goes as follows: 
The time required to drift a distance $x \sim N$ 
is $t \sim N^{1/\mu}$, hence
\beq
v \ \ \sim \ \ \frac{x}{t} \ \ \sim \ \ \left(\frac{1}{N}\right)^{\frac{1}{\mu}-1}
\eeq
This result has meaning only within the Sinai regime ${s<s_1}$.
One observed that the dependence of $v$ on $N$ 
is either sub-Ohmic or super-Ohmic depending 
whether $s$ is smaller or larger than $s_{1/2}$.
Note that for $s\sim0$ the same argument leads to 
the well known Sinai suppression factor $\sim \exp\left(-\sqrt{N}\right)$
that reflects the build-up of an activation barrier.     
}

\begin{figure}
\centering
\includegraphics[height=6.5cm]{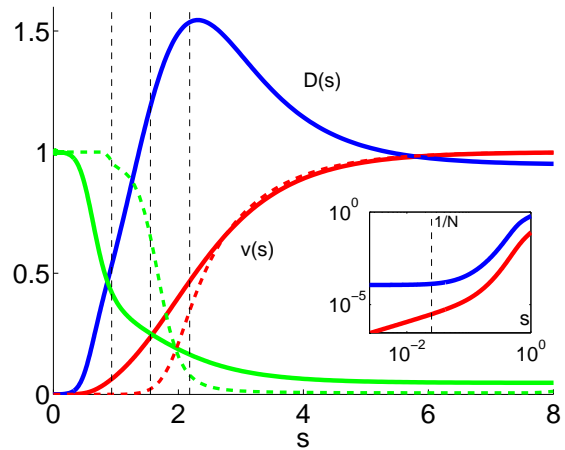}
\caption{
{The drift velocity $v_s=v(s)$ [labeled red lines], 
the monotonic decreasing $a_s/N$, 
and the diffusion coefficient $D=D(s)$ [labeled blue lines] 
as a function of~$s$. 
The time units are chosen such that $v_{\infty}=\langle 1/\ora{w}\rangle^{-1}=1$.
The horizontal lines from left to right correspond \
to $s=1/N$ (inset), and $s_{1/2}$, and $s_1$, and $s_2$. 
They divide the $s$ dependence into 5 distinct regimes (see text). 
Solid lines are for $N=20$ and $\sigma=3.5$.
The dashed lines are for $N=300$. For the latter 
the step in $v$ (at $s=s_1$) and the drop in $a$ (at $s=s_2$) 
become more pronounced.}
}
\label{fig4}
\end{figure}

\begin{table*}
\begin{tabular}{|l||c|c|c|c|c|}
\hline
$s$ regime & $[0, 1/N]$  &  $[1/N, s_{1/2}]$  &  $[s_{1/2}, s_{1}]$  & $[s_{1}, s_{2}]$  & $[s_{2},\infty]$ \\
\hline
$a_s$ & irrelevant 
& \multicolumn{3}{|c|}{$ a_s \sim  N$} 
& $a_s \approx \left[1-\eexp{-2(s-s_{2})}\right]^{-1} \!\!\! a_{\infty}$ \\
\hline
$v_s$ & $v = 2D\,s$ 
&  \multicolumn{2}{|c|}{$\sim \left(\frac{1}{N}\right)^{\frac{1}{\mu}-1}$} 
&  \multicolumn{2}{|c|}{$v_s \approx \left[1-\eexp{-(s-s_1)}\right] v_{\infty}$} \\
\hline
$D$  
& $\sim \exp\left(-\sqrt{N}\right)$ 
& $\sim \left(1/N\right)^{\frac{1}{\mu}-2}$ 
& $\sim \left(N\right)^{2-\frac{1}{\mu}} $
& $\sim N $ 
& $D = \frac{1}{2}a_sv_s$ \\
\hline
\end{tabular}
\caption{Transport characteristics in the various $s$ regimes. 
In the $s<s_1$ regime the transport is suppressed due to Sinai 
activation-barrier that scales with $N$. In the $s>s_1$ regime 
the bias stretches away this barrier, and the drift approaches  
its resistor-network limited value $v_{\infty}$. Normal diffusion is recovered for $s>s_2$.}
\end{table*}

\section{The Sinai anomaly}

For a non-periodic disordered lattice ($N=\infty$) 
it has been found \cite{odh1} that outside 
of the Sinai regime (${s>s_1}$) the the drift velocity is 
\be{90}
v_s \ = \  
\frac{ 1 - \left\langle (\ola{w}/\ora{w})\right\rangle}
{\left\langle (1/\ora{w}) \right\rangle}
\ = \ \left[1-\eexp{-(s-s_1)}\right] v_{\infty}  
\ee
{
The exact expression for the diffusion coefficient is also known 
but is quite lengthy \cite{odh1}. In practice we find that 
we can deduce $D$ from \Eq{e91} if we make the identification  
\be{92}
a_s \ \approx \ \frac{a_{\infty}}{1- \left\langle (\ola{w}/\ora{w})^2\right\rangle}
\ = \ \frac{a_{\infty}}{1-\eexp{-2(s-s_{2})}}
\ee
This expression holds in the normal diffusion regime $s>s_2$, 
where $D$ is finite.  We emphasize again that this expression 
is implied {\em by inspection} of a very complicated formal expression \cite{odh1}
whose physics we would like to illuminate below. In a practical
perspective our approach opens the way for an easy extension 
to finite~$N$, which we present in the next section.  
}

The result of the $v/D$ calculation using \Eq{e91} with \Eq{e92} 
is displayed in \Fig{fig2}. Due to the Sinai anomaly we have ${v=0}$ 
within a finite range $s<s_1$, hence the ESR is completely violated. 
For large but finite $N$ we observe in \Fig{fig2} 
the remnants of the Sinai anomaly, which we call ``Sinai step".    
The question arises what does it mean ``large $N$". 
For this purpose let us use a hand-waving argument 
in order to illuminate the reason for having a vanishingly small 
drift velocity. In a quasi-equilibrium situation 
we have ${ \ora{w}_np_n = \ola{w}_n p_{n{+}1} }$. 
It follows that $p_n \sim \alpha^n$ where 
\beq
\alpha \ \ \equiv \ \ \left\langle \frac{\ola{w}}{\ora{w}} \right\rangle^{-1} 
\ \ = \ \ \eexp{s-s_1}
\eeq      
Transport into the sample is possible provided ${\alpha>1}$, 
else, if ${[\log(1/\alpha)]N \gg 1}$, the probability for penetration 
is blocked. Thus a small $s$ regime with vanishingly small 
drift velocity is feasible if ${N \gg a^{(\sigma)}}$, where  
\be{211}
a^{(\sigma)} \ \ = \ \ \frac{1}{s_1} \ \ = \ \ 
\left[ 
\ln\left( \frac{\sinh\sigma}{\sigma} \right)
\right]^{-1}
\eeq
The dependence of $a^{(\sigma)}$ on $\sigma$ is illustrated 
in \Fig{fig3}. In the numerics we assume large~$\sigma$ values,  
so there are remnants of the Sinai anomaly. 
For weak disorder (${\sigma \ll 1}$) only a very large sample 
will exhibit the ``Sinai step" that we see in \Fig{fig2}.


\section{The $s$ dependence of $D$}

{
The generalized ESR \Eq{e1} with the affinity-dependent length scale $a(s)$
can be used in order to deduce the dependence of $D$ on $s$ 
for a finite $N$ system with disorder. For $s<1/N$ we have 
the conventional linear dependence that is predicted by the traditional ESR. 
For $s>1/N$ the diffusion coefficient $D$ is determined 
by the product $v_s a_s$, see \Eq{e91}. 

Recall that the $N$ dependence of $v_s$ goes from sub-Ohmic to super-Ohmic at $s=s_{1/2}$, 
and becomes $N$ independent for $s>s_1$. 
Recall also that $a_s$ scales like $N$ for $s<s_2$, 
and becomes $N$ independent for $s>s_2$. 
Accordingly as $s$ increases we have the following regimes:
{\bf \ (1)} Linear hopping regime for $s<1/N$.
{\bf \ (2)} Hopping regime with small $D$ for $s \in [1/N, s_{1/2}]$. 
{\bf \ (3)} Hopping regime with large $D$ for $s \in [s_{1/2}, s_1]$.
{\bf \ (4)} Sliding regime with large $D$ for $s \in [s_1, s_2]$.  
{\bf \ (5)} Sliding regime with normal diffusion for $s > s_2$.
We use the keywords ``Hopping" and ``Sliding" in order to 
indicate whether $v_s$ vanishes or not in the $N\rightarrow\infty$ limit.
We use the keywords ``small" and ``large" with regard to $D$ in order to 
indicate whether it vanishes or diverges in the $N\rightarrow\infty$ limit,   
while ``normal" means finite result in this limit. 
The generic $s$ dependence is illustrated in \Fig{fig4} 
and summarized in Table~I.
}

\section{Discussion}
In general non-equilibrium circumstances the bare ESR is not valid. 
For the model system of interest we write its generalized version 
as \Eq{e1} with \Eq{e2}, 
where $a_s$ is an $s$ dependent effective scale
that depends on microscopics of the sample.
It is only in the quasi-equilibrium limit $s\rightarrow0$ 
that the traditional ESR becomes valid.
{As the affinity~$s$ is increased, the length-scale $a_s$ 
drops from the maximal value ${a_0=N}$ (the periodicity)
to the disorder limited value value $a_{\infty}$.
The latter depends monotonically on the strength $\sigma$ of the disorder.} 
 
The first impression is that the generalized ESR \Eq{e1} with \Eq{e2} is very wrong.
Naively the ESR should apply also in non-equilibrium circumstances  
because it can be derived from the NFT assuming CLT. 
We have explained that the resolution of this puzzle is related 
to the implicit coarse-graining procedure.
Consequently the effective affinity is $\overline{s}=f_{\sigma}(s)$.
One wonders what is the "small parameter" on which the ESR is based. 
Considering Brownian motion on an ${a=1}$ periodic lattice 
the answer is that the affinity should be small ($sa\ll1$). 
For a disordered lattice with period $N$ the effective lattice 
constant $a_s$ becomes larger, and hence the condition $sa_s\ll1$ 
becomes more demanding. 

{
The implication of coarse graining is relevant experimentally 
if the resolution of the measurement apparatus is limited \cite{saar}. 
A recent example presented itself in an experimental test 
of the NFT for electron transport through a quantum dot \cite{nftexp1,nftexp2,nftexp3}.
The explanation of the apparent violation of the NFT has been 
based on the elimination of secondary loops in the circuit \cite{gsprd1,gsprd2}, 
hence what counts is not the bare affinity but the effective affinity.
In the present study we have assumed that topological ambiguities 
are absent. There are no secondary loops, just simply connected circuit. 
Still we see that the NFT cannot be applied using the bare affinity.
Replacing it by an effective affinity one obtains the generalized~ESR. } \\

\sect{Acknowledgments}
This research was supported by the Israel Science Foundation (grant No.29/11).
We thank Saar Rahav (Technion) for motivating discussions, 
{and Pierre Gaspard (ULB) for a useful advice.}


\clearpage
\appendix  
\onecolumngrid

\section{Procedure for calculating $v$ and $D$}

We present here the general procedure for calculating $v$ and $D$ 
of a diffusive particle on a lattice that has an $N$~site 
unit cell. For presentation purpose let us consider 
for example a lattice with a 2~site unit cell. 
The rate equation for (say) sites $n=3,4$ takes the form 
\beq
\dot{p}_{3} &=& \ora{A}p_{2} -(\ola{A} + \ora{B})p_{3} + \ola{B}p_{4}\\
\dot{p}_{4} &=& \ora{B}p_{3} -(\ola{B} + \ora{A})p_{4} + \ola{A}p_{5}
\eeq
where $p_n$ are the occupation probabilities of the infinite lattice. 
Applying Bloch theorem the {\em right} eigenvectors are 
determined by two amplitudes $\psi_1$ and $\psi_2$, 
and the recursion $\psi_{n+2} = \eexp{i\varphi}\psi_{n}$, 
where the Bloch phase ${\varphi \equiv kN}$ is used to define 
the quasi-momentum~$k$. The reduced equation for the Bloch amplitudes is  
\beq
\left(
\begin{array}{cc}
 -(\ola{A}+\ora{B}) & \ora{A}\eexp{-i\varphi}+\ola{B}  \\
 \ora{B}+\ola{A}\eexp{i\varphi} & -(\ola{B}+\ora{A}) 
\end{array}
\right) 
\left(
\begin{array}{c}
 \psi_{1}\\
 \psi_{2}
\end{array}
\right) 
\ \ = \ \ 
- \lambda
\left(
\begin{array}{c}
 \psi_{1}\\
 \psi_{2}
\end{array}
\right) \quad\quad
\eeq
The minus sign in front of the eigenvalues is a matter of convention.
Note that for $\varphi=0$ one obtains the lowest eigenvalue ${\lambda_0=1}$ 
which is associated with the NESS. 
Schematically we write the reduced equation as $\bm{W}\psi=-\lambda\psi$.
The generalization for $N$~site unit cell is straightforward.   
Using Dirac notations the reduced Bloch equation is  
\beq
\bm{W}(\varphi) |\varphi,\nu \rangle \ \ = \ \ -\lambda_{\nu}(\varphi) |\varphi,\nu \rangle
\eeq
Since $\bm{W}$ is not a symmetric matrix, 
one should distinguish between left and right eigenvectors. 
The left eigenvectors are defined via the equation 
\beq
\langle \varphi,\tilde{\nu}| \bm{W}(\varphi) \ \ =  \ \ -\lambda_{\nu}(\varphi)\langle \varphi,\tilde{\nu}|
\eeq
Optionally the latter can be regarded as the right eigenvectors of $\bf{W}^{\dag}$
\beq
\bm{W}^{\dag}(\varphi)|\varphi,\tilde{\nu} \rangle \ \ = \ \ -\lambda^{*}_{\nu}(\varphi)|\varphi,\tilde{\nu} \rangle 
\eeq
The left and right eigenvectors form a complete basis
\beq
\sum_{\varphi,\nu} |\varphi,\nu \rangle  \langle \varphi,\tilde{\nu}| &=& \bm{1}\\
\langle \varphi,\tilde{\nu}   |\varphi',\nu' \rangle &=& \delta_{\varphi,\varphi'} \delta_{\nu,\nu'}
\eeq
Turning back to the full lattice, disregarding normalization and gauge issues, 
the Bloch states can be written in the traditional way as a modulated plane wave: 
\beq
\big\langle n \big| k, \nu \big\rangle \ \ = \ \ 
\big\langle n \, \text{mod}(N) \big| \varphi_k, \nu \big\rangle
\ \eexp{ikn}
\eeq   
where $\varphi_k$ is related to $k$ as defined previously.
Consequently the time dependent solution of the rate equation is  
\beq
p_n(t) \ \ = \ \ \sum_{k,\nu}
\eexp{-\lambda_{\nu}(k) t} \  
\langle n|k, \nu \rangle \langle k,\tilde{\nu}| \text{initial-state} \rangle 
\eeq
Averaging the probability within each unit cell, 
we get rid of the intra-cell modulation, leading to  
\beq
p_n(t) \ \ \approx \ \ \frac{1}{L}\sum_{k,\nu} C_{k,\nu} \ \eexp{-\lambda_{\nu}(k) t} \ \eexp{ikn}
\eeq 
where $L$ is the length of the sample, and $C_{k,\nu}$ are constants 
that depend on the initial preparation.
Note that in the limit $k\to 0$ the lower bands degenerate, reflecting a unique NESS. 
Furthermore, due to normalization ${C_{0,0}=C=1}$.
The moment generating function that is associated with $p_n(t)$ is 
\beq
Z(k) \ \ = \ \ \sum_n \eexp{-ikn} p_n(t)  \ \ \approx \ \ 
\sum_{\nu} C_{k,\nu} \ \eexp{-\lambda_{\nu}(k) t} 
\eeq
The first and second moments of $n$ can be deduced by taking the 
first and second derivative of $Z(k)$ at $k=0$. 
In the long time limit, only the $\nu=0$ band survives.
Expanding to second order in $k$ we get
\beq
Z(k) \ \ \approx \ \  
\left[
C + C' k + \frac{1}{2}C''k^2 + ... 
\right]
\left[
1 - \Big(\lambda'k + \frac{1}{2}\lambda''k^2 +...\Big) t 
+ \frac{1}{2} \Big(\lambda'k +...\Big)^2 t^2 + ... 
\right]
\eeq
From which we deduce that in the long time limit
\beq
\langle n \rangle & \approx & i( C' - C \lambda' t ) \\
\langle n^2 \rangle & \approx& -( C'' - C\lambda'' t  - 2C'\lambda' t +  C\lambda'^2 t^2)\\
\text{Var}(n) &\approx & C\lambda'' t - C'' + C'^2 
\eeq
where $C{=}1$, while $\lambda'$ and $\lambda''$ are Taylor coefficients 
in the expansion of $\lambda_0(k)$. The mobility and the diffusion coefficient 
are determined accordingly: 
\be{164}
v \ &=& \ \lim_{t\to\infty} \left[\frac{\langle n \rangle }{t}\right] 
\ \ = \ \ \left. i \frac{\partial \lambda_0(k)}{\partial k}\right|_{k=0}
\\ \label{e165}
D \ &=& \ \lim_{t\to\infty}\left[\frac{1}{2} \frac{\text{Var}(n)}{t}\right]  
\ \ = \ \ \left. \frac{1}{2} \frac{\partial^2 \lambda_0(k)}{\partial k^2}\right|_{k=0}
\eeq

\ \\

\section{Calculation of $v$ and $D$ for $N{=}2$ lattice}

For the two site system, the lowest eigenvalue $\lambda_0(k)$ is 
\beq
\lambda_0(k) 
&=& \frac{1}{2}  
\left[
(\ora{A}+\ola{A}+\ora{B}+\ola{B}) 
- \sqrt{ (\ora{A}+\ola{A}+\ora{B}+\ola{B})^2
-4 (1-\eexp{-i \varphi})\ola{A} \ola{B}
-4 (1-\eexp{i \varphi})\ora{A} \ora{B} }
\right]  
\\
& \approx & 
-i \left[
\frac{\ora{A} \ora{B}-\ola{A} \ola{B}}{\ora{A}+\ola{A}+\ora{B}+\ola{B}}
\right] \varphi   
+ \left[
\frac{\ora{A} \ora{B} + \ola{A} \ola{B}}{2(\ora{A}+\ola{A}+\ora{B}+\ola{B})} 
-\frac{\left(\ora{A} \ora{B}-\ola{A} \ola{B}\right)^2}{\left(\ora{A}+\ola{A}+\ora{B}+\ola{B}\right)^3}
\right] \varphi^2
\eeq
From which the mobility and diffusion coefficients are derived
\beq
v \ &=& \ \frac{\ora{A} \ora{B}-\ola{A} \ola{B}}{\ora{A}+\ola{A}+\ora{B}+\ola{B}} \\
D \ &=& \ \frac{\ora{A} \ora{B}+\ola{A} \ola{B}}{2(\ora{A}+\ola{A}+\ora{B}+\ola{B})}
-\frac{\left(\ora{A} \ora{B}-\ola{A} \ola{B}\right)^2}{\left(\ora{A}+\ola{A}+\ora{B}+\ola{B}\right)^3}
\eeq

\clearpage
\end{document}